\RequirePackage{fix-cm}
\documentclass[smallextended]{article}       
\usepackage{graphicx}
\usepackage{amsmath}
\usepackage{amssymb}
\usepackage{threeparttable}
\usepackage[affil-it]{authblk}
\begin{document}

\title{The MORA project\thanks{The MORA project is supported by R\'egion Normandie. M.G-A. is supported by a Marie Sklodowska-Curie Individual Fellowship of the European Commissions Horizon 2020 Programme under contract number 745954 Tau-SYNERGIES.}
}


\author[1]{P. Delahaye\thanks{corresponding author: pierre.delahaye@ganil.fr}}  
\author[2]{E. Li\'enard}
\author[3]{I. Moore}
\author[2]{M. Benali}
\author[4]{M. L. Bissell} 
\author[3]{L. Canete}
\author[3]{T. Eronen}
\author[5]{A. Falkowski}
\author[2]{X. Fl\'echard}
\author[6]{M. Gonzalez-Alonso}
\author[7]{W. Gins}
\author[3]{R. P. De Groote}
\author[3]{A. Jokinen}
\author[3]{A. Kankainen}
\author[6]{M. Kowalska}
\author[1]{N. Lecesne}
\author[1]{R. Leroy}
\author[2]{Y. Merrer}
\author[6,7]{G. Neyens}
\author[1]{F. De Oliveira Santos}
\author[2]{G. Quemener}
\author[3]{A. De Roubin}
\author[1]{B.-M. Retailleau}
\author[1]{T. Roger}
\author[7]{N. Severijns}
\author[1]{J. C. Thomas}
\author[1]{K. Turzo}
\author[1]{P. Ujic}


\affil[1]{GANIL, CEA/DSM-CNRS/IN2P3, Bd Henri Becquerel, 14000 Caen, France}
\affil[2]{Normandie Univ, ENSICAEN, UNICAEN, CNRS/IN2P3, LPC Caen, 14000 Caen, France}
\affil[3]{University of Jyv\"askyl\"a, department of Physics, 40014 Jyv\"askyl\"a, Finland}
\affil[4]{Nuclear Physics Group, School of Physics and Astronomy, The University of Manchester, Manchester, M13 9PL, United Kingdom}
\affil[5]{Laboratoire de Physique Th\'{e}orique (UMR8627), CNRS, Univ. Paris-Sud, Universit\'{e} Paris-Saclay, 91405 Orsay, France}
\affil[6]{CERN, EP-SME, 1211 Geneva, Switzerland}
\affil[7]{KU Leuven, Instituut voor Kern- en Stralingsfysica, 3001 Heverlee, Belgium}

\date{
Created: \today 
 }

\maketitle

\begin{abstract}
The MORA (\textbf{M}atter's \textbf{O}rigin from the \textbf{R}adio\textbf{A}ctivity of trapped and oriented ions) project aims at measuring with unprecedented precision the $D$ correlation in the nuclear beta decay of trapped and oriented ions. The $D$ correlation offers the possibility to search for new CP-violating interactions, complementary to searches done at the LHC and with electric dipole moments. Technically, MORA uses an innovative in-trap orientation method which combines the high trapping efficiency of a transparent Paul trap with laser orientation techniques. The trapping, detection, and laser setups are under development. The project will first focus on the proof-of-principle of the in-trap laser orientation technique, before the actual measurement of the $D$ correlation in the decay of $^{23}$Mg$^+$ ions is undertaken firstly at JYFL and then later, at GANIL, with full sensitivity to new physics.
\end{abstract}
\section{Introduction}
\label{intro}
Why are we living in a world of matter? What is the reason for the strong matter – antimatter asymmetry we observe in the Universe?

The MORA (\textbf{M}atter's \textbf{O}rigin from the \textbf{R}adio\textbf{A}ctivity of trapped and oriented ions) project aims at searching for possible hints. In 1967, A. Sakharov expressed the 3 conditions which should be fulfilled for the baryogenesis process \cite{5}. These conditions are: (i) a large C and CP violation; (ii) a violation of the baryonic number, (iii) a process out of thermal equilibrium. A large CP violation has therefore to be discovered to account for this large matter-antimatter asymmetry, at a level beyond the CP violation predicted to occur in the Standard Model via the quark-mixing mechanism.

\begin{figure}
\begin{center}
  \includegraphics[width=11cm]{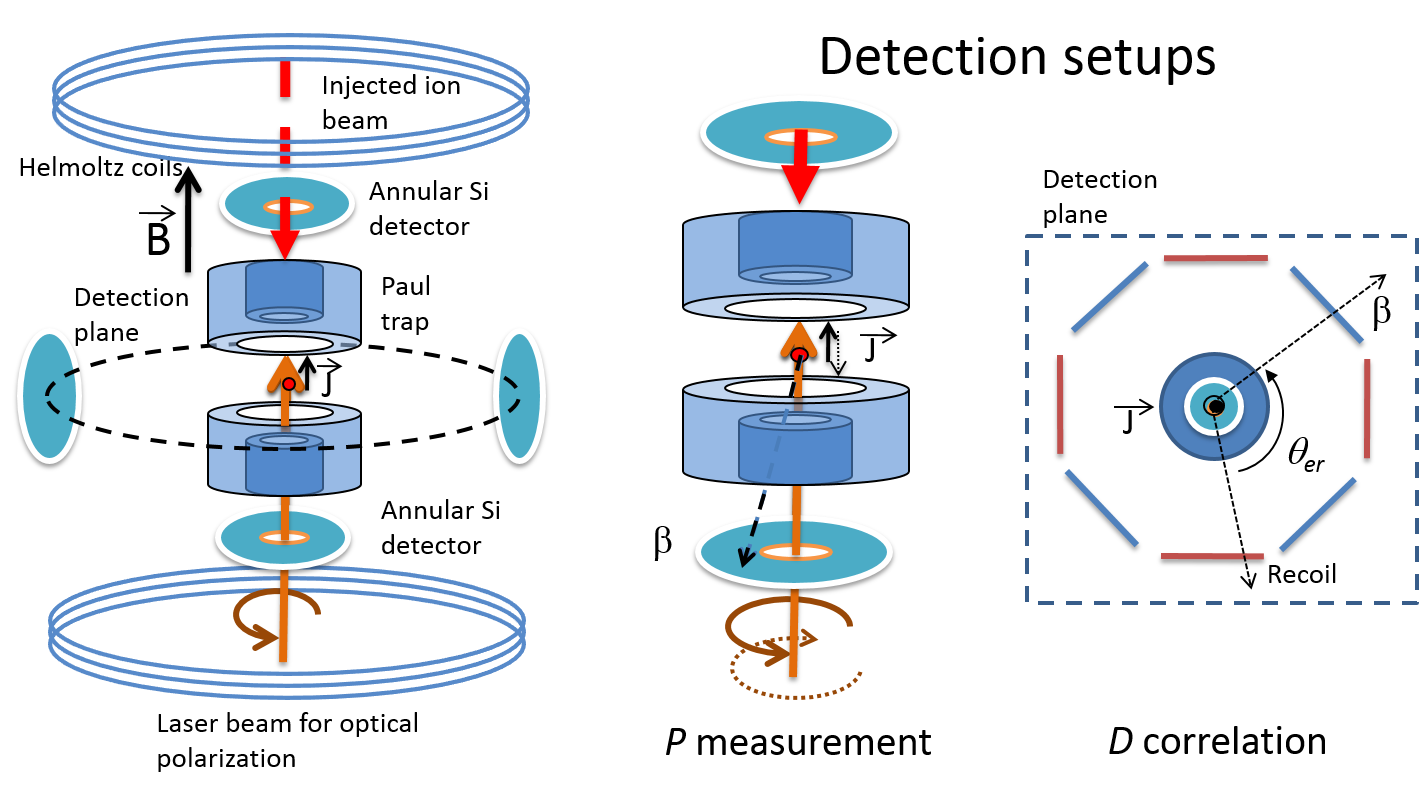}
\end{center}
\caption{In trap optical polarization and detection setup. The polarization is obtained by superimposing the laser beam with the injected ion beam. Helmholtz coils maintain a preferential axis for the B-field in the trap. The Helmholtz coils used to cancel transverse stray $B$ fields in the azimuthal plane are not shown for more visibility. The polarization is monitored by observing the count rate difference for positrons detected in the two annular Si detectors on the axis of the trap. The $D$ correlation is measured in the azimuthal plane of the trap by an emiT – like \cite{28} arrangement of beta and recoil ion detectors, consisting respectively of phoswhich and Micro-Channel Plate (MCP) detectors.}
\label{fig:ovvw}       
\end{figure}

MORA aims at measuring the $D$ correlation \cite{33} in the nuclear beta decay of trapped and oriented ions with unprecedented precision. There is a large window in which $D$ correlation and Electric Dipole Moment (EDM) measurements can contribute to the search for other sources of CP violation at a much higher level than predicted by the Standard Model \cite{6}.  The $D$ correlation offers the possibility to search for new CP-violating interactions in a region that is less accessible by EDM searches, in particular via the Leptoquark model \cite{7,8,9,34}. Leptoquarks are heavy bosons coupling leptons and quarks which appear in many extensions of the Standard Model. Their decay out of thermal equilibrium in the framework of Grand Unified Theories came rather evidently as the first hypothesis for the baryon and lepton number non conservation \cite{11,12,13}. As such, Leptoquarks still play a peculiar role among the many models now formulated for baryogenesis. Remarkably, they were lately invoked to explain recent lepton flavor non conservation in the decay of B mesons \cite{10}. With sensitivity on $D$ close to 10$^{-5}$, the MORA apparatus should additionally permit a probe of the Final State Interactions (FSI) effects for the first time. The latter can mimic a non-zero $D$ correlation of the order of 10$^{-5}$ to 10$^{-4}$, depending on the beta decay transition which is observed. 
Technically, MORA uses an innovative in-trap orientation method which combines the high trapping efficiency of a transparent Paul trap with laser orientation techniques. The MORA setup is schematically shown in Fig. \ref{fig:ovvw}. The project will first focus on the proof-of-principle of the in-trap laser orientation technique, before the measurement of the $D$ correlation in the decay of $^{23}$Mg$^+$ ions is undertaken firstly at JYFL and then later, at GANIL, with full sensitivity to new physics.
\section{Correlation measurements in $\beta$ decay}
\label{sec:cor}
The measurement of correlations appearing in the nuclear beta decay spectrum provides stringent constraints on non standard model interactions (see e.g. \cite{3,4} and refs. therein). The competitiveness of such measurements with respect to LHC searches for new interactions has recently been investigated (see e.g. \cite{3,4,54}) using an Effective Field Theory (EFT) framework, which encompasses a wide class of New Physics (NP) models. Bounds for exotic vector, axial-vector, tensor and scalar interactions were compared. For some of the nonstandard interactions LHC searches provide limits that are competitive with (or sometimes much stronger than) beta decays. In contrast, the LHC cannot offer competitive bounds for the specific CP-violating interaction affecting the $D$ correlation \cite{54,50}, which thus presents an excellent NP potential.
For future reference, the correlations appearing in the decay of oriented nuclei are listed in Tab. \ref{tab:1} together with their dependence on the kinematics parameters of the decay and their intrinsic symmetries: $m_e$, $p_e$, $E_e$ are the electron mass, momentum, and total energy, $p_\nu$, $E_\nu$ the momentum and energy of the neutrino, and $J$ the spin of the nucleus. P stands for the parity transformation, T for time reversal. The correlations with capital letters refer to nuclei of oriented spin, while the Fierz interference term $b$ and the $\beta-\nu$ angular correlation parameter $a_{\beta\nu}$ are also encountered in decays of non-oriented nuclei. Other types of correlations appear when measuring the polarization of electrons or positrons, for example the triple R correlation, of the form $R\vec{\sigma}\cdot\frac{\left< \vec{J} \right>}{J}\times\frac{\vec{p_e}}{E_e}$.  
LPCTrap presently permits a measurement of the beta neutrino angular correlation $a_{\beta\nu}$, which does not require any polarization, to an aimed precision of 0.5\%. In principle, the MORA trapping and detection setup will enable the measurement of the $a_{\beta\nu}$, $A_{\beta}$, $B_{\nu}$ correlations, in addition to $D$. Although not considered in the framework of MORA for the search of NP, these correlations can provide important information on the parameters of the ion cloud. In particular, we intend to monitor the polarization degree via the measurement of the $A_{\beta}$ and $B_{\nu}$ correlations, as developed in section \ref{sec:pol}.
\begin{table}
\caption{Correlations appearing in the $\beta$ decay for oriented nuclei \cite{33}.  See text for details.}
\label{tab:1}   
\begin{tabular}{lll}
\hline\noalign{\smallskip}
Name of the correlation & Correlation term & Symmetry \\ 
\noalign{\smallskip}\hline\noalign{\smallskip}
Beta neutrino angular correlation & $a_{\beta\nu} \cdot \left( \frac{\vec{p_e}}{E_e}\cdot \frac{\vec{p_\nu}}{E_\nu} \right)$ & P even T even \\  
Fierz interference term & $b\cdot \left( \frac{m_e}{E_e} \right)$ & P even T even \\  
Beta asymmetry  & $A_\beta\cdot \left( \frac{\left< \vec{J} \right>}{J}\cdot \frac{\vec{p_e}}{E_e} \right)$ & P odd \_ T even \\ 
Neutrino asymmetry & $B_\nu\cdot \left( \frac{\left< \vec{J} \right>}{J}\cdot \frac{\vec{p_\nu}}{E_\nu} \right)$ & P odd \_  T even \\ 
$D$ correlation & $D\cdot \left( \frac{\left< \vec{J} \right>}{J}\cdot \left(\frac{\vec{p_e}}{E_e}\times \frac{\vec{p_\nu}}{E_\nu} \right) \right)$ & P even T odd \\ 
\noalign{\smallskip}\hline
\end{tabular} 
\end{table}
\section{The MORA apparatus}
\label{sec:app}
As shown in Fig. \ref{fig:ovvw}, the trap of MORA will be installed into a dedicated chamber with ports for the laser beam transport. The trap and detection configuration will allow for the simultaneous measurement of the $D$ correlation and monitoring of the degree of polarization.

\begin{figure}
\begin{center}
\includegraphics[width=8cm]{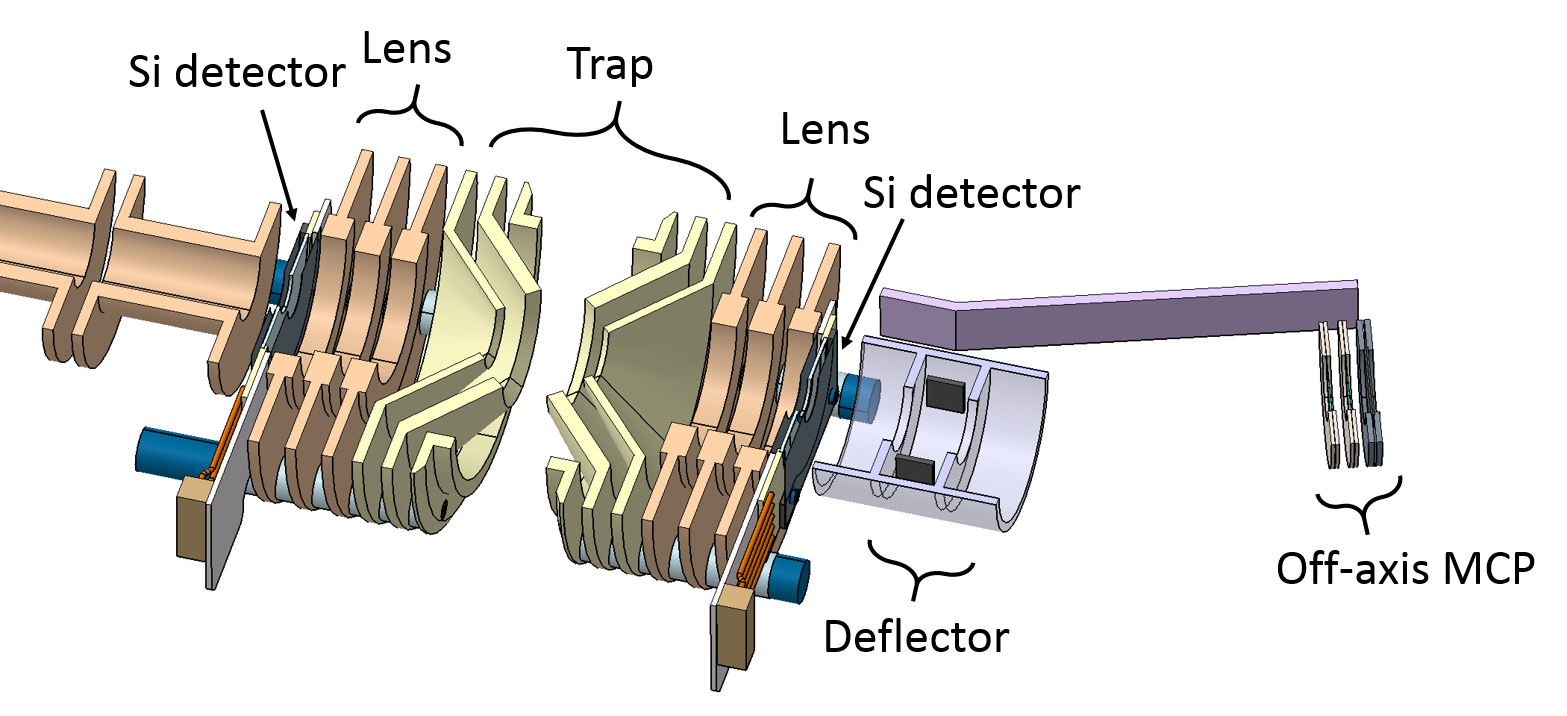}
\end{center}
\caption{Cross section of the trap electrodes and of the surrounding elements. The two annular Si detectors monitor the degree of polarization of ions. An off-axis MCP permits an evaluation of the number of ions trapped cycle per cycle, after ejection of the ion cloud towards the deflector electrodes.}
\label{fig:trap}       
\end{figure}

Originally inspired from LPCTrap \cite{35}, the trap design has been optimized for enlarging the quadrupolar region, where higher order harmonics are negligible. The conical shapes of the electrodes, visible in Fig. \ref{fig:trap}, offer a slightly larger solid angle for detection. The trap optimization will be described in an upcoming article. Using the latter geometry and a more powerful RF, enhanced trapping performances are expected compared to LPCTrap. One aims at increasing both the space charge capacity and the trapping lifetime by one order of magnitude by using optimized RF power. The present limits in LPCTrap correspond to a space charge capacity of about $5\cdot 10^5$ ions per cycle \cite{30} and a lifetime of $\sim$ 0.5 s due to evaporative collisions from the shallow trap potential \cite{35}. In such a trap, an overall ion beam cooling, injection and trapping efficiency of the order of few tens of \% can reasonably be achieved, provided that the trap is not filled to its maximum space charge capacity.  A typical measurement cycle will consist of the injection of a fresh bunch of ions into the trap; a short preparation period for reaching a saturated polarization degree ($\ll$100ms, see Sec. \ref{sec:pol}); and a measurement period, the duration of which will depend on the ion trapping lifetime and radioactive decay half-life. For $^{23}$Mg, a typical cycle will last a few seconds. 

\begin{figure}
\begin{center}
\includegraphics[width=10cm]{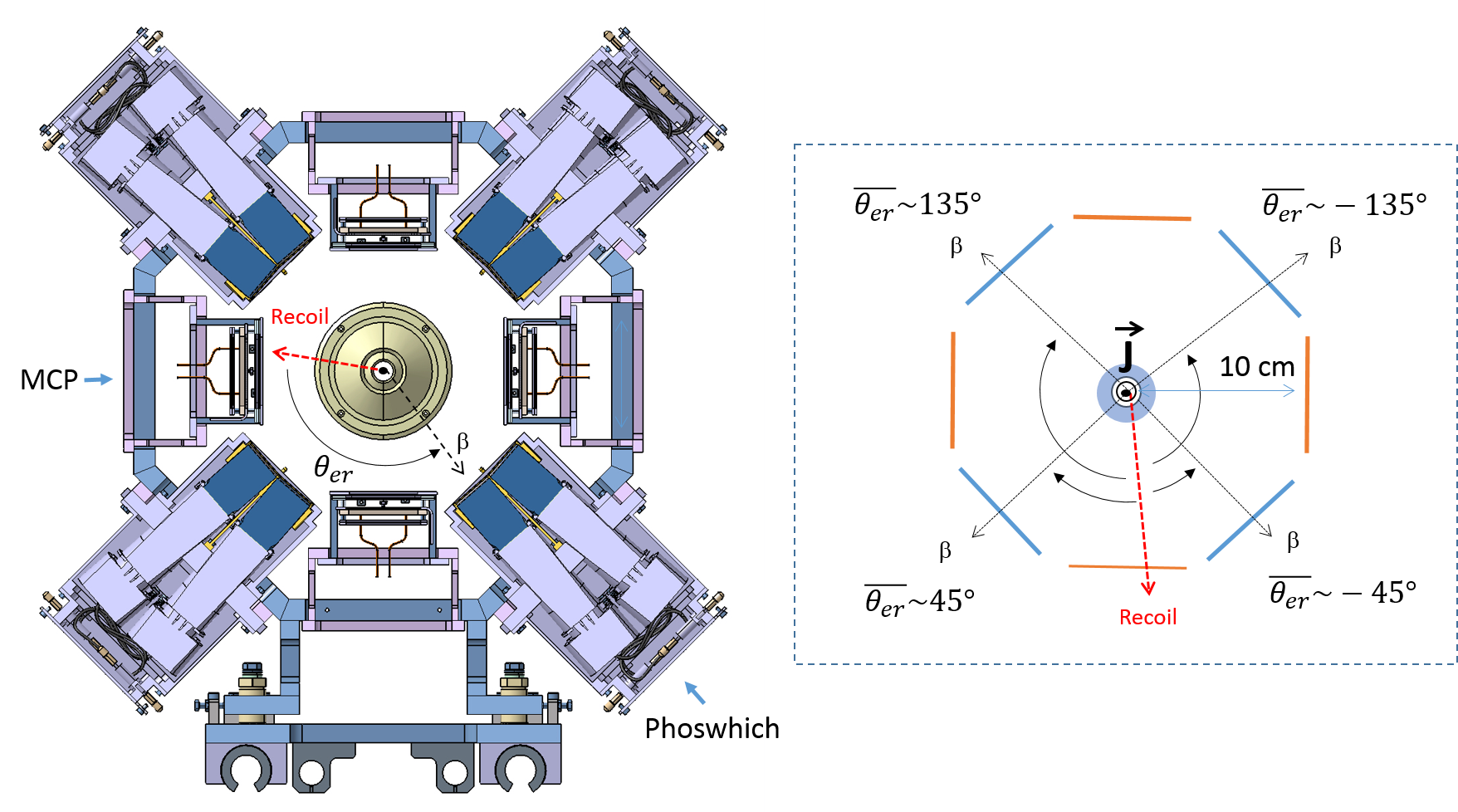}
\end{center}
\caption{Left inset: cross section of the $D$ correlation detection setup. The trap electrodes are visible in the center of the detection setup. Right inset: sketch showing the different types of $\beta$-recoil coincidences which will be recorded by the detection setup. }
\label{fig:detD}       
\end{figure}

\begin{figure}
\begin{center}
\includegraphics[width=10cm]{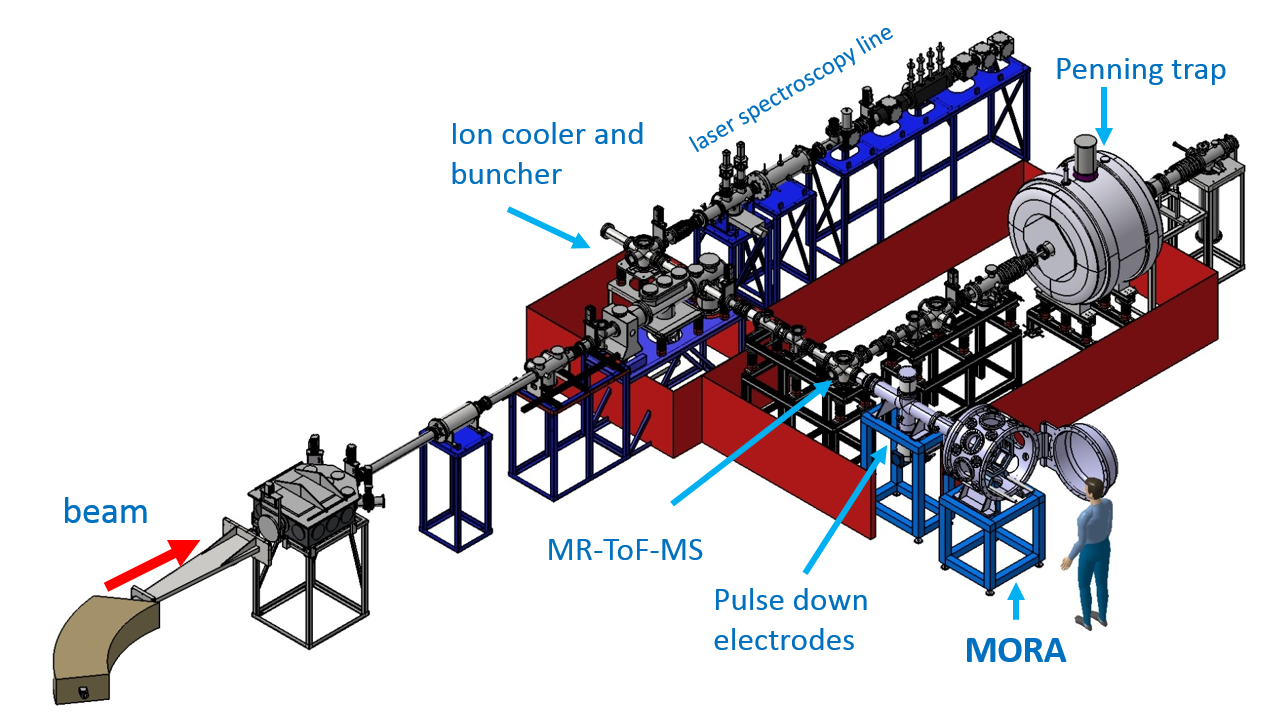}
\end{center}
\caption{Installation of the MORA apparatus in the IGISOL-4 beamlines. }
\label{fig:IG}       
\end{figure}

On the axis of the trap, two segmented annular detectors from Micron will monitor the polarization degree (see Fig. \ref{fig:trap} and Sec. \ref{sec:pol}). The $D$ correlation will be measured in the azimuthal plane of the trap. The detection configuration presented in Fig. \ref{fig:detD}, enables a very good sensitivity to the correlation (see Sec. \ref{sec:meas}). It is very similar to the one used in the emiT experiment \cite{28}. The left inset shows the 4-fold segmented phoswich detectors, which will be detecting the $\beta$ particles, and the Micro-Channel Plates (MCPs) with accelerating grids and a home made position-sensitive anode which will detect recoiling ions with a nominal 50\% efficiency. The phoswich detectors are being presently being tested, while the final designs of the MCP and Si detectors are discussed with Photonis and Micron respectively. The MORA apparatus will be shipped and installed in the IGISOL-4 beamlines before the fall of 2020 (see Fig. \ref{fig:IG}).

\section{Laser polarization of the $^{23}$Mg$^+$ ions}
\label{sec:pol}
So far, trap-based experiments aiming at measuring observables with polarized nuclei have been achieved with atoms using Magneto-Optical-Traps (MOTs), sometimes in combination with other types of atom traps, see e.g. \cite{17} and \cite{18}. In \cite{17} the degree of polarization is determined from the photo-ionization of atoms in a non-polarized state, while in \cite{18} a sophisticated method of re-trapping atoms in a MOT was used. In both setups a degree of polarization above 99\% was achieved and determined with a precision in the per-mille range.  In contrast to radioactive atom trapping in MOTs, radioactive ion trapping can be an extremely efficient process (more than 10\%), especially since the advent of ion cooling techniques in RFQ cooler bunchers \cite{19}.
 The degree of polarization can be expected to be equally high (close to 100\%) because of the extended exposure of the confined ions to the laser light. The proof-of-principle of the laser polarization will be performed at JYFL at the IGISOL-4 facility \cite{20}, with a pulsed laser system. The outlined combination of an ion trap and laser polarization is a new concept, taking advantage of the high trapping efficiency of the transparent Paul trap and the high degree of polarization from optical pumping. This method has never been used for such correlation measurements in nuclear $\beta$ decay, although laser pumping has recently been applied in a RFQ cooler-buncher for collinear laser spectroscopy \cite{21}, and in linear Paul traps, where ion laser cooling allows the study of quantum phase transitions in crystals of a few trapped ions or the development of new generation atomic clocks \cite{22}.
  
\begin{figure}
\begin{minipage}[c]{.46\linewidth}
      \includegraphics[width=5.7cm]{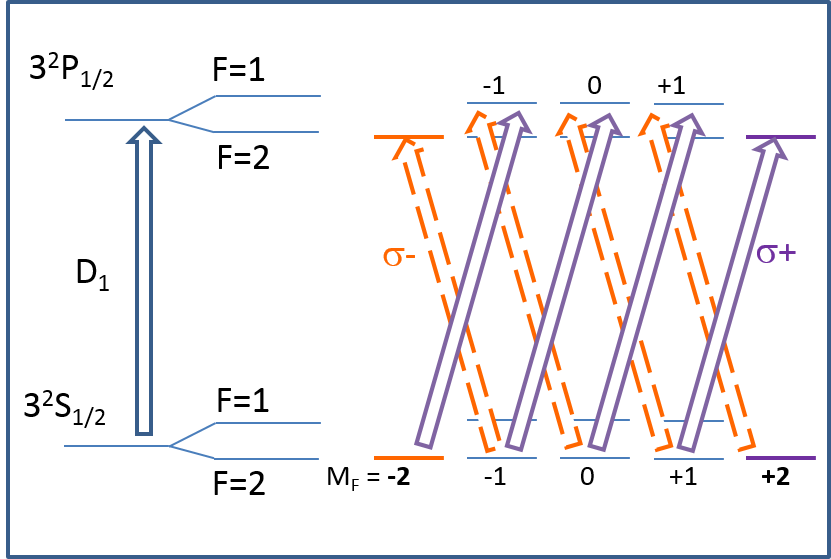}
   \end{minipage} \hfill
   \begin{minipage}[c]{.46\linewidth}
      \includegraphics[width=5.5cm]{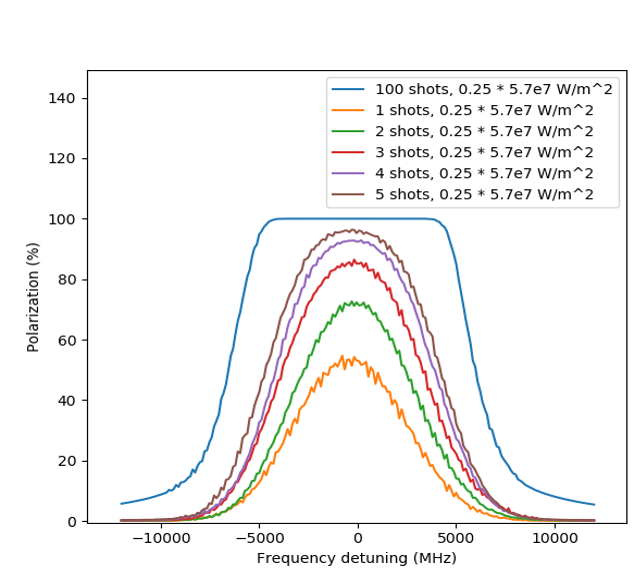}
   \end{minipage}
\caption{Left: D1 hyperfine transition used to orient the spin of $^{23}$Mg$^+$ ions ($\sim$280 nm). F=I+J  is the sum of the nuclear spin I and of the electron cloud J. The polarization can be easily reverted during the experiment by changing the laser circular polarization from $\sigma+$ to $\sigma-$. Right: polarization efficiencies for a varying number of 5 $\mu$J pulses spread over an ion cloud of diameter of 3 mm. }
\label{fig:pol}       
\end{figure}

We will first focus on $^{23}$Mg$^+$, which is a suitable candidate for both laser pumping and $\beta$ decay correlation measurements. 
$^{39}$Ca$^+$ will be considered at a later stage. A RFQ cooler buncher is readily available in the IGISOL-4 beamlines \cite{20} and will provide bunched and cooled beams of $^{23}$Mg for efficient trapping in the MORA apparatus. The optical pumping schemes to polarize Mg ions are well established \cite{24,25,26}. 
Recent simulations indicate that the optimal laser system will consist of a single broadband pulsed Ti:Sa laser, whose frequency will be tripled to cover the transitions leading to the maximum magnetic substates M$_F=\pm2$ shown in the left inset of Fig. \ref{fig:pol}. Such a laser system is readily available at IGISOL-4.  The laser beam will be circularly polarized using a linear polarizer and quarter wave plate, and guided via suitable transport optics to the trap. The simulations take into account the Doppler effect due to the thermal motion of trapped ions. Assuming a typical repetition rate of 10 kHz and 20 $\mu$J per pulse (0.2 W average power) the achievable nuclear polarization degree after 1 ms (10 pulses) of light exposure would essentially be limited by the degree to which the laser light could be circularly polarized, saturating at a level above 99\%. Similar results are obtained after 10 ms for a lower power of 5 $\mu$J per pulse (see right inset of Fig. \ref{fig:pol}). 
Ions are cooled via collisions with the He buffer gas, at a typical pressure of 10$^{-5}$ mbar. Owing to the expected low collision rate with the most abundant $^4$He atoms (kHz), for which no depolarization in collisions with the ground state (S=I=J=0) is expected, possible depolarization due to collisions with the residual gas between the pulses will be marginal. 
Special care will therefore be taken to monitor the degree of polarization of the laser light by standard optical methods. The polarization will be easily inverted by changing the light circular polarization from $\sigma+$ to $\sigma-$. Helmholtz coils placed outside the vacuum chamber will maintain a B field of the order of a fraction of a mT directed along the axis of the trap to prevent depolarization. 
The polarization degree will be determined by measuring an asymmetry in the counts recorded by two annular $\beta$ detectors placed along and against the axis of the trap as shown in the middle inset of Figure \ref{fig:ovvw}, taking the form:
\begin{equation}
\label{eq:1}
\frac{N_{\beta^+}^\uparrow-N_{\beta^+}^\downarrow}{N_{\beta^+}^\uparrow+N_{\beta^+}^\downarrow}\propto A_\beta\cdot P
\end{equation}
where $N_{\beta^+}^\uparrow$ and $N_{\beta^+}^\downarrow$ are the numbers of positrons detected parallel and anti-parallel to the polarization direction, respectively, and $P$ the polarization degree.
The Standard Model value of $A_\beta$ is tabulated for some of the mirror isotopes in \cite{27} and is known to 0.3\% for $^{23}$Mg: $A_\beta=-0.5584 \pm 0.0017$.
 The use of coincidences with recoils measured in the azimuthal plane of the trap, with the recoil ion detectors serving also for the $D$ correlation measurement (right inset of Fig. \ref{fig:ovvw} and section \ref{sec:meas}) will a priori be preferred to single $\beta$ events, as coincidences permit a high rejection of background \cite{51}. The coincidence with the recoil ions will induce constraints on the neutrino momentum of the detected events, because of the law of momentum conservation. In the case of $\beta$'s conditioned by the coincident detection of a recoil ion, the asymmetry of Eq. \ref{eq:1} can be reformulated as a linear combination of the $\beta$ and neutrino asymmetry coefficients:
\begin{equation}
\label{eq:2}
\frac{N_{\beta^+ \ coinc}^\uparrow-N_{\beta^+ \ coinc}^\downarrow}{N_{\beta^+ \ coinc}^\uparrow+N_{\beta^+ \ coinc}^\downarrow}= \left( \alpha \cdot A_\beta + \beta B_\nu \right) \cdot P \ .
\end{equation}
Here $N_{\beta^+ \ coinc}^\uparrow$ and $N_{\beta^+ \ coinc}^\downarrow$ are the numbers of coincidences between recoil ions detected in the azimuthal plane of the trap, and positrons detected parallel and anti-parallel to the polarization axis in the two annular detectors, and $\alpha$ and $\beta$ are coefficients depending on the detection solid angle and decay parameters. These coefficients can be precisely determined by a Monte Carlo simulation, and $B_\nu$ is known to 0.54\% precision: $B_\nu=-0.7404\pm 0.0040$ \cite{27}. With this method the precision aimed at in the determination of the degree of polarization will be of the order of 2\%, which is more than sufficient for looking for a non-zero D correlation. The systematic uncertainties arising in the measurement of $P$ from non-uniform detection efficiencies will be efficiently reduced by regularly inverting the direction of polarization.  Eq. (2) shows that in first approximation the relative accuracy on $P$ will scale as the square root of the number of $\beta$ – recoil coincidences recorded in the different detectors:
\begin{equation}
\label{eq:3}
\frac{\sigma_P}{P} \simeq \frac{\sqrt{N_{\beta^+ \ coinc}^\uparrow+N_{\beta^+ \ coinc}^\downarrow}}{N_{\beta^+ \ coinc}^\uparrow-N_{\beta^+ \ coinc}^\downarrow} \ .
\end{equation}
Other optical methods (e.g. monitoring the fluorescence signal induced by a laser probe) for the precise determination of the degree of polarization will additionally be investigated. An independent measurement of the degree of polarization would for instance enable the precise measurement of the $A_\beta$  and $B_\nu$  correlations (see section \ref{sec:cor}), which could then be used for other tests of the Standard Model, for example the determination of $V_{ud}$ from mirror transitions \cite{4}.
\section{$D$ correlation measurement}
\label{sec:meas}
\begin{figure}
\begin{center}
\includegraphics[width=8cm]{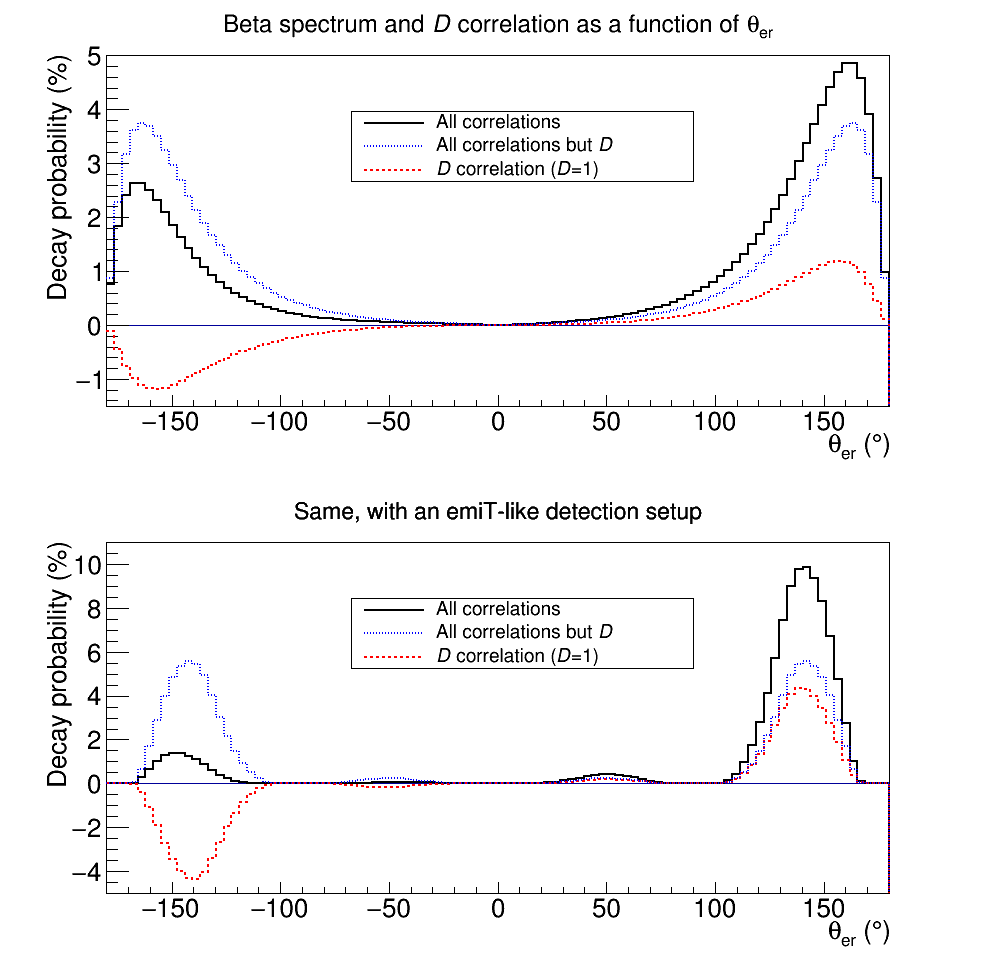}
\end{center}
\caption{Contribution of the $D$ correlation to the $\beta$ decay spectrum of $^{23}$Mg$^+$ ion. $\theta_{er}$ is the angle between the electron and recoil in the plane perpendicular to the polarization axis. The $D$ coefficient is arbitrarily magnified to 1 for more visibility. Black line: integral $\beta$ decay spectrum as a function of $\theta_{er}$, which contains all additional relevant correlations: a$_{\beta\nu}$, A$_\beta$ and B$_\nu$ from \cite{27}. Blue dotted line: same, without the contribution of the $D$ correlation. Red dotted line: contribution of the $D$ correlation. Upper inset: no phase space cut. Bottom inset: same but filtered by the detection setup as shown in Fig. \ref{fig:ovvw} and \ref{fig:detD}.  }
\label{fig:cor}       
\end{figure}

The $D$ correlation will be maximal in the detection plane of MORA shown in Fig. \ref{fig:ovvw}, the product $\vec{p_e}\times \vec{p_\nu}$ being equivalent to $\vec{p_e}\times \vec{p_r}$, where $\vec{p_r}$ is the ion recoil momentum, because of momentum conservation.  The magnitude of the correlation as a function of the angle between the electron and the recoil $\theta_{er}$ in the plane perpendicular to the polarization axis is shown in the upper inset of Fig. \ref{fig:cor}.
It is maximal for an angle approaching 165$^{\circ}$, similar to the neutron decay. The detector arrangement will therefore be close to the one of the emiT experiment \cite{28}, alternating every 45$^{\circ}$ a detector for the recoil ion and a detector for the $\beta$ particle, as shown in Fig. \ref{fig:detD}. The lower inset of Fig. \ref{fig:cor} shows the $D$ correlation as would be observed by such a detection setup, assuming an octagonal arrangement of 6 cm diameter recoil ion and positron detectors placed at 10 cm from the trap center. Given the polarization direction, the $D$ correlation can be inferred from an asymmetry in the number of coincidences recorded at average $\theta_{er}$ angles of +45$^{\circ}$, +135$^{\circ}$ on one side and -45$^{\circ}$, -135$^{\circ}$ on the other side, the sign of $\theta_{er}$ being defined clockwise with respect to the spin direction (see Fig. \ref{fig:detD}).
\begin{equation}
\label{eq:4}
\frac{N_{coinc}^{+45^{\circ}}+ N_{coinc}^{+135^{\circ}}-N_{coinc}^{-45^{\circ}}- N_{coinc}^{-135^{\circ}}}{N_{coinc}^{+45^{\circ}}+ N_{coinc}^{+135^{\circ}}+N_{coinc}^{-45^{\circ}}+ N_{coinc}^{-135^{\circ}} }=\delta \cdot D \cdot P \ .
\end{equation}
$N_{coinc}^{+45^{\circ}}$, $N_{coinc}^{+135^{\circ}}$, $N_{coinc}^{-45^{\circ}}$, $N_{coinc}^{-135^{\circ}}$ refer to the number of recorded coincidences at a given average $\bar{\theta}_{er}$, and $\delta$ is a constant depending on the detection solid angle and decay parameters. 

$\delta$ can be precisely determined by a Monte Carlo simulation of the $\beta$ decay spectrum and detection setup. In the conditions assumed in this first simulation (point-like ion cloud, no RF disturbance of the recoil, no detection threshold) it is equal to $0.775(1)$ and has to be compared to the factor $\left| K \right| \sim 0.378$ for the emiT experiment quoted in \cite{15}. 
A more realistic simulation will not give a large difference in the $\delta$ value (not more than a few percent), as the listed approximations correspond to second order effects: the size of the ion cloud is of the order of 1 mm of radius in comparison to $\sim$3 cm radius detectors.
The RF disturbance can be corrected by recording the RF phase as presently done for the measurement of $a_{\beta\nu}$ \cite{51}. 
As is presently the case for the $a_{\beta\nu}$ measurement, low detection thresholds will be achievable for both the electron detectors and the MCPs detecting the recoiling ions, which will be equipped with grids to post-accelerate the ions from a few tens of eV to a several keV to reach the efficiency plateau \cite{29}.
 The sizeable gain of sensitivity compared to the emiT experiment is attributed to the selection of the plane of the decay, perpendicular to the polarization axis, which becomes possible because of the confinement of the emitting source (ions in our case) in a small volume in the center of the trap.  Systematic uncertainties arising from non-uniform detection efficiencies and polarization will be reduced by regularly inverting the direction of polarization, and using the symmetries of the octagonal arrangement.  In \cite{15} combinations of asymmetries of neighbouring detectors additionally permit the suppression of unwanted contributions of other correlations. Background and electron backscattering related effects will be determined in the kinematically forbidden regions as readily done in \cite{51}.
 In experiments aiming at setting limits on a non-zero $D$ correlation, the precise knowledge of $P$ is obviously not required, as long as $P$ can be maintained over 70-80\%. 
  $P$ will be monitored during the $D$ correlation measurement using coincidences of the recoiling ions with $\beta$’s detected in the annular Si detectors, as described in section \ref{sec:pol}. The sensitivity on a non - zero $D$ correlation (eq. \ref{eq:4}) will scale as:
\begin{equation}
\label{eq:5}
 D \simeq \left( \delta\cdot P \cdot \sqrt{N_{coinc}^{+45^{\circ}}+ N_{coinc}^{+135^{\circ}}+N_{coinc}^{-45^{\circ}}+ N_{coinc}^{-135^{\circ}}} \right) ^{-1}\ .
\end{equation}
Despite its apparent simplicity Eq. \ref{eq:5} is in agreement with the statistical considerations given in e.g. \cite{28}. Using equations \ref{eq:3} and \ref{eq:5}, and the results of the Monte Carlo simulation of the $\beta$ spectrum, the relative accuracy on $P$ and sensitivity to a non - zero $D$ can be estimated for the different phases of MORA.  Table \ref{tab:2} summarizes the results of such estimates, where the first column corresponds to the expected number of trapped ions per cycle at JYFL and GANIL. A cycle time of 2 s was considered.

At SPIRAL 1, the expected rates are of the order of $5\cdot 10^8$ pps for $^{23}$Mg. DESIR, which includes a High Resolution Separator and a number of purification traps, will be the ideal place for providing a clean beam for MORA. It will host a laser setup (“LUMIERE”) presently missing in the low energy beam lines of GANIL for the polarization of $^{23}$Mg$^+$ ions. In these conditions, the number of trapped ions per cycles will be mostly limited by the capacity of the trap. 

In the case of JYFL, the statistics may be limited by different factors. The first one is the $^{23}$Mg beam intensity. The second one is the amount of $^{23}$Na contamination. A marginal contamination of the shorter-lived $^{23}$Al is additionally expected, although it should not harm the polarization and $D$ correlation measurement. 

Two competitive reactions were recently compared for MORA in IGISOL-4: $^{nat}$Mg(p,d)$^{23}$Mg, with 30 MeV p, and $^{23}$Na(p,n)$^{23}$Mg with 10 MeV p. For both reactions, about $4\cdot 10^4$ pps of $^{23}$Mg could be produced with 2 $\mu$A of p, while for safe operation, the primary proton beam would be essentially limited to 10 $\mu$A. With a $^{23}$Na target, the second reaction resulted in a level of contamination from $^{23}$Na at least one order of magnitude higher than for the first one: a few $10^8$ pps compared to $10^7$ pps, respectively. 
Even by choosing the appropriate target, the level of contamination would be very difficult to handle in the IGISOL RFQ beam cooler used to prepare the $^{23}$Mg bunches prior to their injection into the transparent Paul trap of MORA. With a typical capacity of $5\cdot 10^5$ ions/bunch, the number of $^{23}$Mg injected in the trap would be limited to about $10^3$ ions per cycle. We plan to reduce the contamination by using a new dedicated RF sextupole ion guide for the extraction of beams from IGISOL, with the aim to trap $2\times 10^4$ $^{23}$Mg$^+$ ions per cycle.  An optimized RF amplifier circuit as used for the high intensity RFQ beam cooler of LPC Caen \cite{53} will additionally increase the IGISOL RFQ beam cooler capacity. If found insufficient, the use of the future MR-ToF-MS of IGISOL will be considered.

The estimates of column 3 and 4 of Tab. \ref{tab:2} are based on eq. \ref{eq:3} and \ref{eq:5}, respectively. The total number of coincidences was deduced from the number of trapped ions given in column 1 and the detection solid angle given in Sec. \ref{sec:app}. An average MCP detection efficiency of 25\% was additionally assumed, accounting for the  $\sim$50\% probability for the recoils to remain neutral after the $\beta$ decay of $^{23}$Mg$^+$ ions, while the others will be ionized by electron shake-off \cite{52} and detected with a typical 50\% efficiency. Column 3 and 4 only take into account the statistical uncertainty. The latter was found to be dominant in all $D$ measurement carried out so far. As can be seen, three weeks of beam time, which could be organized in successive campaigns at DESIR, should be sufficient to reach a sensitivity well below $10^{-4}$ for the $D$ correlation, even considering a very conservative trapping capacity.

%
%
%
%
\begin{table}
\caption{Statistical considerations for the measurements proposed at JYFL/IGISOL-4 and GANIL/DESIR. The statistical uncertainty on the degree of polarization $P$ and the sensitivity on the  $D$ correlation in the decay of $^{23}$Mg$^+$ ions are calculated using eq. \ref{eq:3} and \ref{eq:5}, respectively.}
\begin{center}
\begin{threeparttable}
\label{tab:2}  
\begin{tabular}{p{2.2cm}p{1.5cm}p{1.7cm}p{1.5cm}p{1.5cm}}
\hline 
Place and type of measurement & Trapped ions / cycle & Measur. duration (d) & $\frac{\sigma_P}{P}$ stat. \% & Sensivity on $D$ \\ 
\hline 
JYFL: $P$ & $2\cdot 10^4$ & 8 & $1.9$ & $1 \cdot 10^{-3}$ \\ 
JYFL: $D$ & $2\cdot 10^4$ & 32 & $9.4 \cdot 10^{-1}$ & $5.2 \cdot 10^{-4}$ \\
DESIR: $D$\tnote{1} & $1\cdot 10^6$ & 24 & $1.5 \cdot 10^{-1}$ & $8.5 \cdot 10^{-5}$ \\
DESIR: $D$\tnote{2} & $5\cdot 10^6$  & 24 & $6.9 \cdot 10^{-2}$ & $3.8 \cdot 10^{-5}$ \\
\hline 
\end{tabular} 
\begin{tablenotes}
\item [1] With a conservative trapping capacity.
\item [2] With a nominal trapping capacity.
\end{tablenotes}
\end{threeparttable} 
\end{center}
\end{table}
\section{Sensitivity to New Physics}
\label{sec:sens}
In beta decay, the triple correlations can only be motion-reversal odd. As such, the triple $D$ and $R$ correlations are, under CPT invariance, sensitive to new sources of CP violation \cite{31}.
At higher energy, CP violation has been observed in the decay of the kaon, B, and most recently D0 mesons. CP violation is incorporated in the Standard Model via the quark mixing mechanism, but at a level which cannot account for the large matter – antimatter asymmetry observed in the Universe. $D$ and $R$ are sensitive probes which offer a large window to search for a much larger CP violation. This window will eventually be limited by the level at which the effects of electromagnetic FSI, which mimic the existence of non-zero $D$ and $R$ correlations, can be calculated \cite{8}. The FSI effects have never been observed so far at the present level of precision on $D$ and $R$. Such effects are typically of the order of $10^{-4}$ to $10^{-5}$ depending on the system, and have been recently calculated to an absolute accuracy of $10^{-7}$ for the neutron \cite{32}. $D_{FSI}$ values of the order of $1.3\cdot 10^{-4}$ and $-0.47\cdot 10^{-4}$ can be estimated using for instance \cite{14} for $^{23}$Mg and $^{39}$Ca, respectively.
The bounds on imaginary T-violating scalar and tensor interactions are obtained from measurements of the $R$ correlation in the decay of the neutron and $^8$Li, respectively \cite{3,4}. Stronger bounds are obtained from the LHC using the same analysis as for the real parts of these exotic interactions, in such a way that an improvement in future $R$ measurements will be necessary to keep them competitive. 
The situation is quite different for the $D$ correlation, which sets bounds on a CP violating V, A interaction which are weakly constrained by LHC data \cite{54,50}. 
The best constraints so far on such T-violating interaction arise from the neutron decay, with a measurement yielding $D_n=(-0.9 \pm 2.1 \cdot 10^{-4})$ \cite{15,55}.
 Weaker constraints were obtained for nuclear decays using the decay of $^{19}$Ne, with a combination of measurements yielding a limit $D_{^{19}Ne}<6\cdot 10^{-4}$ \cite{16}. 
In terms of the coupling constants from the effective Hamiltonian of beta decay, the present limits on $D_n$ translate into a limit on the relative phase between the vector and axial couplings: Im$(C_V/C_A )=(1.6\pm 6.3)\cdot 10^{-4}$ (90\% C.L.) \cite{3}. 
In practice the neutron and nuclear $\beta$ decays have a different sensitivity to the relative phase between the vector and axial couplings, which would bring evidence for NP.
This sensitivity can be expressed as $D_X =F(X) \cdot$Im$(C_V/C_A )$, where $X$ is the neutron or $\beta$ decaying nucleus, and $F(X)$ is a factor which depends on the nuclear spin and Gamow Teller to Fermi ratio \cite{33}. 
$F(X)$ is displayed in Tab. \ref{tab:3} for the neutron and different isotopes. The sensitivity of $D$ to NP is found to be superior for some nuclear $\beta$ decays, by 50\% in the case of $^{23}$Mg and $^{39}$Ca, which can be in-trap polarized in the MORA setup.
\begin{table}
\begin{center}
\caption{Sensitivity $F(X)$ of the $D$ correlation to NP for the decay of different $X$ initial states.}
\label{tab:3}   
\begin{tabular}{lllll}
\hline 
Particle $X$ & n & $^{19}$Ne & $^{23}$Mg & $^{39}$Ca \\ 
\hline 
$F(X)$	& -0.55 & 0.66 & 0.82 & -0.90 \\
\hline 
\end{tabular} 
\end{center}
\end{table}
The measurement of the $D$ correlation using the MORA setup presents additional advantages thanks to the confinement of ions: 
\begin{itemize}
\item A 2-fold increase in sensitivity to the $D$ correlation by the selection of the most sensitive plane of detection (see $\delta$ factor in Eq. \ref{eq:4} in section \ref{sec:meas} and the subsequent discussion).
\item A strong reduction of the systematic effects related to the source size, geometry and inhomogeneity of polarization due to the small size, symmetry of the ion cloud and very high polarization degree. In particular, the so-called asymmetric-beam-transverse-polarization and beam expansion which were identified for the neutron as dominant sources of uncertainties in Tab. 4 of ref. \cite{15} will be strongly reduced in the MORA experiment.
\end{itemize}
EDMs in nuclei or in the neutron are other sensitive probes to CP violation. In full generality, these searches are complementary and they can only be compared within specific NP scenarios. In a general EFT framework it is easily shown that they access different combinations of parameters. For some leptoquark models that are not covered by the EFT approach current $D$ measurements of neutron beta decay represent the strongest constraints, even taking into account nEDM bounds \cite{9}. Moreover, the $D$ correlation is theoretically a much cleaner probe to NP than EDM, which relies on hadronic and nuclear computations \cite{8,34}. 

In summary, the development in the coming years of the in-trap polarization of radioactive ion beams with the MORA setup will permit to attain new frontiers of sensitivity to NP, searching for possible sources of CP violation. With a precision down to a few $10^{-5}$, the sensitivity to the imaginary phase Im$(C_V/C_A)$ will be improved by one order of magnitude compared to the present bounds. The FSI effects mimic a non-zero $D$ correlation at a level which is below the precision of the current experiments. Such effects will eventually be probed for the first time, if no signature of NP is found at a higher level, resulting in a stringent test of these calculations. We will first focus on $^{23}$Mg$^+$ which is easier to polarize. This isotope is well produced at JYFL for performing the proof-of-principle of the in-trap polarization, and the first $D$ correlation measurement. This first phase at JYFL (2020-2023) will permit a study of all the parameters that affect the sensitivity of the MORA setup, before the measurements of the $D$ correlation with highest sensitivity to NP are performed at DESIR, where $^{23}$Mg and $^{39}$Ca will be available with copious intensities from the SPIRAL upgrade and S3-LEB facilities (from 2024 onwards).

%
%




\end{document}